\documentclass{PoS}

\newcommand \bra[1]{\left< {#1} \,\right\vert}
\newcommand \ket[1]{\left\vert\, {#1} \, \right>}

\newcommand{\bea}{\begin{eqnarray}}
\newcommand{\eea}{\end{eqnarray}}
\newcommand{\simgt}{\hbox{ \raise3pt\hbox to 0pt{$>$}\raise-3pt\hbox{$\sim$} }}
\newcommand{\simlt}{\hbox{ \raise3pt\hbox to 0pt{$<$}\raise-3pt\hbox{$\sim$} }}

\title{\begin{center}
Computation of Heavy Quarkonium Spectrum\\ in Perturbative QCD
\end{center}
}

\ShortTitle{Computation of Heavy Quarkonium Spectrum in Perturbative QCD}

\author{\speaker{Yukinari Sumino}\\ \\
        Department of Physics, Tohoku University,
Sendai, 980--8578 Japan
}

\abstract{
Non-relativistic bound state theories for QED and QCD have become fairly mature 
and amenable to a textbook-level understanding and computation.
In this talk we give an introductory review of
the following subjects related to the recent computation
of the heavy quarkonium spectrum using perturbative QCD:
(1)~Technological developments in higher-order computation, 
(2) Physics predictions, (3)
Challenge towards analytic evaluation of the 3-loop static QCD potential.
}

\FullConference{Loops and Legs in Quantum Field Theory\\
		24-29 April 2016\\
		Leipzig, Germany}

\begin{document}

\section{Introduction}

In recent years there has been remarkable technological progress in the computation of
higher-order corrections to various high-energy processes.
The main driving force has been the tough demands from the current
LHC experiments, where vast amount of (difficult) computation is indispensable
for extracting information on important physics quests.
Indeed many contributions in this direction have been presented in this workshop.

In this talk we are concerned with a slightly different subject, the
higher-order computation of the heavy quarkonium spectrum.
The motivation to deal with this physical system is as follows.
Heavy quarkonium (such as $t\bar{t}$, $b\bar{b}$, $c\bar{c}$ and $b\bar{c}$)
is unique among various strongly interacting systems, 
in the sense that properties of individual hadrons
can be predicted purely within perturbative QCD.
Observables such as its spectrum, decay width, or level transition rates
have been computed and tested against
experimental data or compared with lattice QCD computations.
Through such procedure we can test perturbative QCD under
clean environment and gain deeper understanding on predictability and
proper usage of perturbative QCD in relation to non-perturbative
effects.
At the same time we can determine the fundamental parameters of
the standard model, such as the heavy quark masses and the strong 
coupling constant, with high accuracy.
(See \cite{Brambilla:2004wf}
and references therein.)

In modern computation of
higher-order corrections to observables of heavy quarkonium,
two theoretical foundations play crucial roles.
One is the effective field theory (EFT) framework,
such as potential-NRQCD (pNRQCD)  \cite{Pineda:1997bj}
or velocity-NRQCD (vNRQCD) \cite{Luke:1999kz}.
The other is the computational technology
called threshold expansion technique \cite{Beneke:1997zp}.
These theoretical tools enable 
organization of computations of higher-order corrections
in a systematic manner.

In this review we explain,
to those who have interests in higher-order computations
in general
but are non-experts of bound-state physics,
the following subjects
related to the recent computation
of the heavy quarkonium spectrum in perturbative QCD:
In Sec.~2,~Recent technological developments in higher-order computation;
In Sec.~3, Physics predictions and applications;
In Sec.~4, Current challenge towards analytic 
evaluation of the 3-loop static QCD potential.
Summary is given in Sec.~5.

\section{Computation of quarkonium spectrum up to NNNLO}

In this section we explain some aspects of
the recent technological developments
in the computation of the heavy quarkonium energy levels.
The state-of-the-art computation is at the next-to-next-to-next-to-leading
order (NNNLO) level \cite{Beneke:2005hg,Kiyo:2014uca,Peset:2015vvi}.
The calculation uses pNRQCD EFT
for systematically organizing the perturbative expansions
in $\alpha_s$.

This EFT describes interactions of a non-relativistic
quantum mechanical system (dictated by the Schr\"odinger equation)
with ultrasoft gluons, which is organized in multipole expansion.\footnote{
This is in analogy to classical electrodynamics, in which electric field with
a long wave-length 
(compared to the scale of charge distribution) can be expressed generally as
a superposition of electric field generated by electric multipoles.
}
Hence, the Lagrangian of pNRQCD is given as expansions in $\vec{r}$ and
$1/m$ in the following form:
\bea
{\cal L}_{\rm pNRQCD}=S^\dagger 
\bigl( i\partial_t - \hat{H}_S \bigr) S +
{O}^{a\dagger}
\bigl(iD_t-\hat{H}_{O} \bigr)^{ab}
{O}^b
+ g \, S^\dagger \, \vec{r}\cdot\vec{E}^a\, {O}^a
+\dots .
\label{pNRQCD-Lagrangian}
\eea
$S$ and ${O}^a$ represent, respectively,
color-singlet and octet heavy quark-antiquark ($Q\bar{Q}$) 
composite fields.
$E^a$ denotes the color-electric field.
$\hat{H}_S$ and $\hat{H}_{O}$ denote
the quantum mechanical Hamiltonians
for the singlet and octet states, respectively, 
and
dictate the main binding dynamics of the $Q\bar{Q}$ states.
(The zeroth-order Lagrangian in expansion in $\vec{r}$
just gives the Schr\"odinger equations for $S$ and $O^a$ as the
equation of motion.)
$\hat{H}_S$ is currently known up to NNNLO \cite{Kniehl:2002br} 
in non-relativistic expansion
(that is, double expansion in $1/m$ and $\alpha_s$).

The energy levels of the heavy quarkonium states are given as
the positions of poles of the full propagator of the
singlet field $S$ in pNRQCD.
\begin{figure}
\begin{center}
\includegraphics[width=9cm]{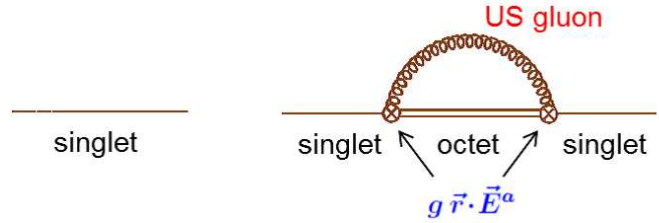}
\end{center}
\vspace*{-.5cm}
\caption{\small
First two diagrams for the full propagator of $S$
in multipole expansion in pNRQCD.
\label{Diagrams-pNRQCD}}
\end{figure}
The full propagator, in multipole expansion in $\vec{r}$,
is given by the diagrams in Fig.~\ref{Diagrams-pNRQCD}.
The first diagram represents the zeroth-order propagator
of $S$, which is simply $1/[E-\hat{H}_S+i0]$.
Hence, its pole positions can be computed by ordinary perturbation theory
of quantum mechanics using $\hat{H}_S$ up to NNNLO.
The LO Hamiltonian is that of the Coulomb system. 
The perturbative corrections are given by the familiar formula:
\bea
\delta E_n=\bra{n}(\hat{H}_S)_{\rm NLO}\ket{n}+
\sum_{i\neq n}
\frac{
|\bra{n}(\hat{H}_S)_{\rm NLO}\ket{i}|^2
}{E_n-E_i}
+\cdots\,.
\eea
Each term of the
perturbative corrections can be straightforwardly
converted to an infinite sum form using a known infinite
sum representation of the Coulomb Green function.

The second diagram represents emission and reabsorption of
an ultrasoft gluon via dipole interaction in eq.~(\ref{pNRQCD-Lagrangian}).
This contribution is the same as Lamb shift in QED and 
contains UV divergences.
The divergences are canceled by IR divergences contained in the Hamiltonian
at NNNLO, $(\hat{H}_S)_{\rm N^3LO}$, such that the
sum of the two diagrams is finite.
We now explain key aspects in the evaluation of
contribution of each diagram in Fig.~\ref{Diagrams-pNRQCD}.

\subsection{Breakdown of infinite sum to finite sums (and known transcendental numbers)
}

Let us explain the new technology to evaluate infinite sums
in the evaluation of the first diagram.
We take the following sum as an example:
\bea
&&
A(n,\ell)=
\sum_{k=1}^{\infty}
\frac{(n-\ell+k-1)!}{(n+\ell+k)! \, k^3} .
\label{A}
\eea
This sum appears as a part of the NNLO corrections to the
spectrum. 
We have devised a new method to reduce this type of infinite sums to
a combination of transcendental numbers 
[such as $\zeta(2)=\pi^2/6$, $\zeta(3)$, etc.], rational numbers and
a finite sum  \cite{Kiyo:2014uca}.

By partial fractioning, the product over the magnetic quantum number $m$
can be written as
\bea
&&
\frac{(n-\ell+k-1)!}{(n+\ell+k)!}=\prod_{m=-\ell}^{\ell}\frac{1}{n+k+m}
=\sum_{m=-\ell}^{\ell}\frac{R(\ell,m)}{n+k+m},
\label{partialfrac}
\eea
where the residue is given by
\bea
R(\ell,m)=\frac{(-1)^{\ell-m}}{(\ell+m)!(\ell-m)!}
\, .
\eea
Now we can reduce the sum eq.~(\ref{A}) using eq.~(\ref{partialfrac}) and
\bea
\sum_{k=1}^\infty \frac{1}{(k+i)\,k^3}=\frac{\zeta(3)}{i}-\frac{\zeta(2)}{i^2}+\frac{S_1(i)}{i^3}
\, .
\label{UseOfWa}
\eea
[$S_1(i)=\sum_{k=1}^i\frac{1}{k}$
denotes the harmonic sum.]

To evaluate\footnote{
A {Mathematica} package ${\it ``Wa"}$, which implements
this summation algorithm, is available
at 
{\bf http://
www.tuhep.phys.tohoku.ac.jp/$\sim$program/} 
with examples and instructions.}
sums such as the one in eq~(\ref{UseOfWa}),
there exists a general algorithm \cite{Anzai:2012xw}
(see also \cite{Blumlein:2010zv})
which can evaluate, e.g.,
\bea
f(i)=\sum_{k=1}^\infty\sum_{m=1}^\infty\frac{(-1)^{m-k}}{
(k+i)^2(k+m+i)(m+2i+1)
}
\label{f(i)}
\eea
by reducing it to a combination of nested sums
\bea
Z\left(n_{\rm max};\{b_{j}\};\{\lambda_{j}\}\right) =
\!\!\!\!\!
\!\!\!\!\!
\!\!\!\!\!
\sum_{n_{\rm max}> n_{1}>n_{2}>\cdots>n_{N}>0}\frac{\lambda_{1}^{n_{1}}\lambda_{2}^{n_{2}}\cdots\lambda_{N}^{n_{N}}}{n_{1}^{b_{1}}n_{2}^{b_{2}}\cdots n_{N}^{b_{N}}},
~~~~~~~
b_j\in \mathbb{N},~
\lambda_j\in \mbox{roots of unity},
\label{nestedsum}
\eea
where $n_{\rm max}=i$ or $\infty$.
In fact, as a factor in the denomanator on the RHS of eq.~(\ref{f(i)}) 
any linear polynomial of internal and external indices
can appear,
while any roots of unity can appear in the numerator.
The algorithm utilizes the fact that the summand of $f(i)$ can be
brought to a form which is invariant under the
shift of the indices,
$i\to i+\Delta i$, $k\to k+\Delta k$, $m\to m+\Delta m$.
Then it is easy to see that, by taking a difference equation of $f(i)$,
the bulk of the sum gets canceled and only ``surface terms'' with
one summation less remain.
By repeating this procedure recursively and summing back, one
can render $f(i)$ to a combination of nested sums, which
can be evaluated in terms of known transcendental numbers,
rational numbers and finite sums.

\subsection{Algebraic derivation of US correction (QCD Bethe log)}

We show how to compute the second diagram 
of Fig.~\ref{Diagrams-pNRQCD} algebraically \cite{Kiyo:2014uca}.
The diagram corresponds to the 
one-loop self-energy of the singlet field $S$ and 
is given by
\bea
E^{us}_{n\ell}=-ig^2 \mu^{2\epsilon}
\frac{T_F}{N_C}\int_0^\infty dt 
\left\langle \vec{r}\cdot\vec{E}^a(t,\vec{0})\,
\exp\left[ -i(\hat{H}_O^{(d)} - E_{n,C}^{(d)})t \right]
\vec{r}\cdot\vec{E}^a(0,\vec{0})
\right\rangle_{n\ell}
.
\label{DefEUS}
\eea
$\langle \cdots \rangle_{n\ell}$ denotes the expectation value
taken with respect to the (external) energy eigenstate $(n,\ell)$ of the singlet
Hamiltonian $H_S^{(d)}$ in $d$ dimensions.
We employ the dimensional
regularization with $d=D-1=3-2\epsilon$.
$E_{n,C}^{(d)}=-\frac{C_F^2\alpha_s^2}{4n^2}m + {\cal O}(\epsilon)$ 
denotes the (leading-order) energy eigenvalue of $H_S^{(d)}$.
We need to keep 
$\epsilon=(3-d)/2$ non-zero until we extract the UV divergences
of $E^{us}_{n\ell}$ explicitly.

The correlation function of the color electric
field can be evaluated using the gluon propagator of the
ordinary Feynman rules as
\bea
&&
\left\langle E^{ia}(t,\vec{0})E^{ja}(0,\vec{0})
\right\rangle
=-i\delta^{aa}\int \frac{d^Dk}{(2\pi)^D}\,
\frac{e^{ik_0t}}{k^2+i0}\,\left(k^ik^j-k_0^2\delta^{ij}\right)
~+~{\cal O}(\alpha_s) .
\eea
After integrating over $t$ 
we obtain
\bea
&&
E^{us}_{n\ell}=\frac{1}{2}C_F g^2 \mu^{2\epsilon}
\, \frac{1-d}{d}\, C(d)\,
\left\langle r^i \left(\hat{H}_O^{(d)} - E_{n,C}^{(d)}\right)^d
r^i \right\rangle_{n\ell}
.
\label{EUS2}
\eea
$C(d)$ is expressed in terms of Gamma functions
and includes $1/\epsilon$ pole due to the UV divergence of
the one-loop integral.
Therefore, we need the other factors up to order $\epsilon$.

We may expand
$(\hat{H}_O^{(d)}-E_{n,C}^{(d)})^{3-2\epsilon}\approx (\hat{H}_O^{(d)}-E_{n,C}^{(d)})^{3}
[1-2\epsilon\log(\hat{H}_O^{(3)}-E_{n,C}^{(3)})]$ and write
\bea
&&
\mu^{2\epsilon}
\left\langle r^i \left(\hat{H}_O^{(d)} - E_{n,C}^{(d)}\right)^d
r^i \right\rangle_{n\ell}
= \left\langle
X - 2\epsilon\, r^i \left(\hat{H}_O^{(3)} - E_{n,C}^{(3)}\right)^3
\log \left(\frac{\hat{H}_O^{(3)} - E_{n,C}^{(3)}}{\mu}\right)
r^i \right\rangle_{n\ell}
\nonumber \\
&&
~~~~~~~~~~~~~~~~~~~~
~~~~~~~~~~~~~~~~~
~~~~~~~~~~~
+ {\cal O}(\epsilon^2)
\eea
with
\bea
\textstyle
X=r^i \bigl(\hat{H}_O^{(d)}\bigr)^3 r^i- \frac{3}{2}
\left\{ \hat{H}_S^{(d)}, r^i \bigl(\hat{H}_O^{(d)}\bigr)^2 r^i \right\}
+ \frac{3}{2}
\left\{ \bigl(\hat{H}_S^{(d)}\bigr)^2, r^i \hat{H}_O^{(d)} r^i \right\}
- \frac{1}{2}
\left\{ \bigl(\hat{H}_S^{(d)}\bigr)^3, \vec{r}^{\, 2} \right\}
,
\label{defX}
\eea
where we have replaced $E_{n,C}^{(d)}$ by the singlet Hamiltonian
inside the expectation value,
taking into account ordering of the operators.
We can then use the commutation relation $[r_i,p_j]=i\delta_{ij}$
and $\bra{n}[\hat{H}^{(d)}_S,{\cal O}]\ket{n}=0$, which hold for
general $d$, to simplify $X$,  and we obtain
an operator $\hat{H}^{us}$, in which the
$1/\epsilon$ terms and finite terms are explicitly separated:
\bea
&&
E^{us}_{n\ell}=
\left\langle \hat{H}^{us} \right\rangle_{n\ell}
-\frac{2C_F\alpha_s}{3\pi}\left\langle
r^i \left(H_O^{(3)} - E_{n,C}^{(3)}\right)^3
\log \left(\frac{H_O^{(3)} - E_{n,C}^{(3)}}{{\mu}}\right)
r^i \right\rangle_{n\ell}
.
\label{EUS3}
\eea
The $1/\epsilon$ part of $\hat{H}^{us}$ exactly cancels the
$1/\epsilon$ part of $(\hat{H}_S)_{\rm N^3LO}$.
The second term on the RHS represents the QCD analogue of the
``Bethe logarithm'' in Lamb shift.

This is the algebraic derivation of the second diagram, and
the result agrees with the previously known result.
The original derivation \cite{Kniehl:2002br} was based on diagrammatic analyses,
which requires manipulation of vertices and propagators
among a set of Feynman diagrams,
using the equation of motion, etc.
In general the algebraic derivation would be more tractable
for non-experts.

\section{Physics predictions}

\begin{figure}
\begin{center}
\includegraphics[width=7cm]{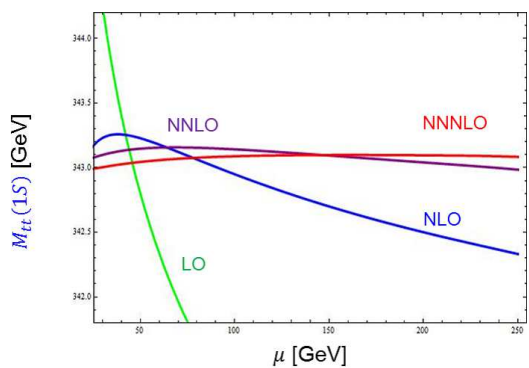}
\end{center}
\vspace*{-.5cm}
\caption{\small
Renormalization-scale dependence of the energy level of
the $t\bar{t}(1S)$ state.
\label{ScaleDepMtt(1S)}}
\end{figure}

The scale dependence of the energy level of the lowest-lying  
spin-one quarkonium state is shown in Fig.~\ref{ScaleDepMtt(1S)}.
The displayed figure is the case of the (would-be) $t\bar{t}(1S)$ state, 
and in the cases of the 
other heavy quarkonium $1S$ states the dependences are qualitatively similar.
As can be seen, better stability is obtained as we include 
higher-order corrections.
One can also verify that the
perturbative series is converging fairly well
around the scale where the $\mu$ dependence
becomes flat (minimal-sensitivity scale).
It is crucial to use a short-distance mass of the
heavy quark to realize these nice features, and
in particular the $\overline{\rm MS}$ mass
shows the best stability and convergence \cite{Kiyo:2015ooa}.
(In this regard,
the recent computation of the four-loop pole-$\overline{\rm MS}$ mass
relation \cite{Marquard:2015qpa} is highly appreciated.)

\begin{figure}[t]
\begin{center}
\begin{tabular}{cc}
\hspace*{-5mm}
\includegraphics[width=7.5cm]{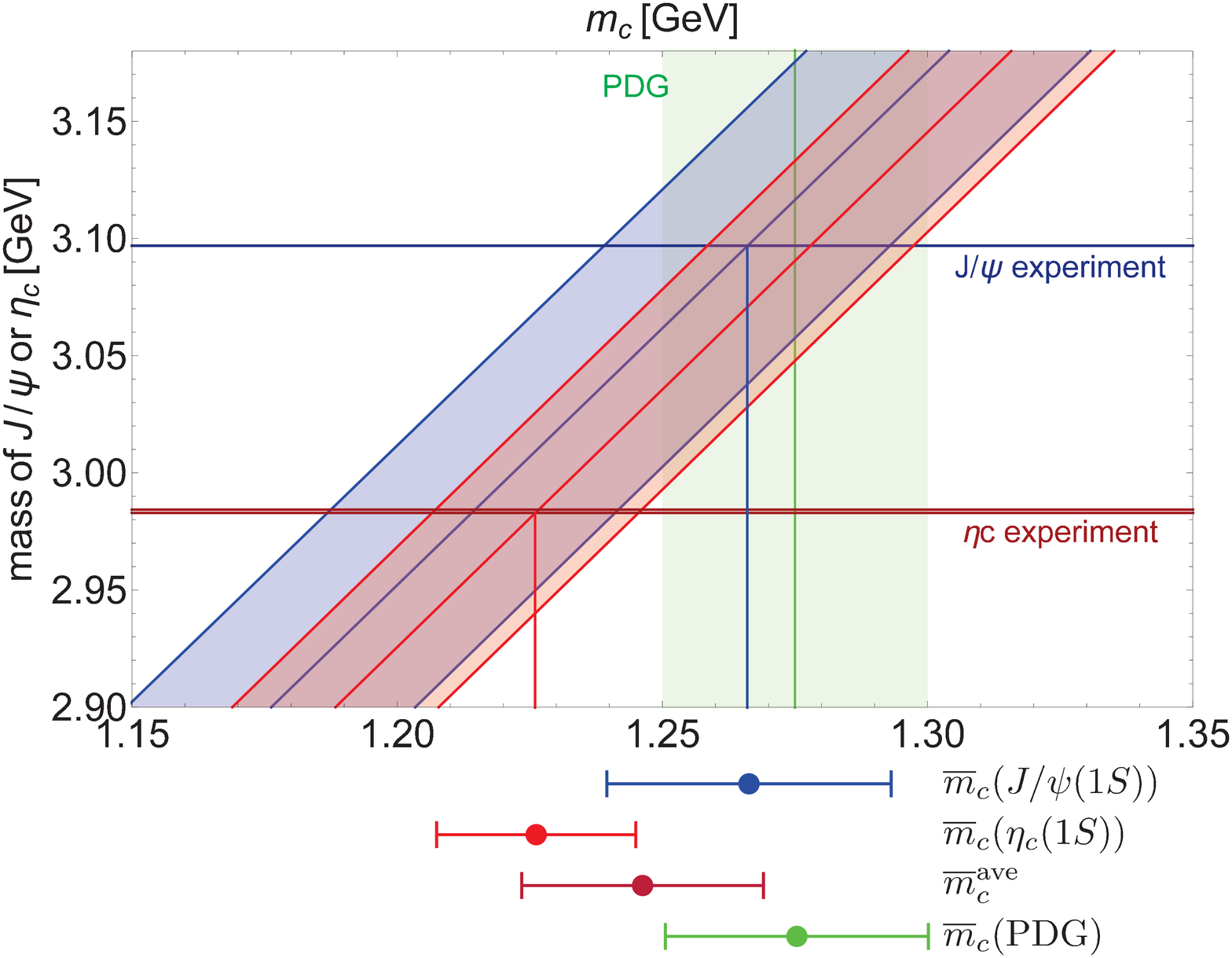}&
\includegraphics[width=7.5cm]{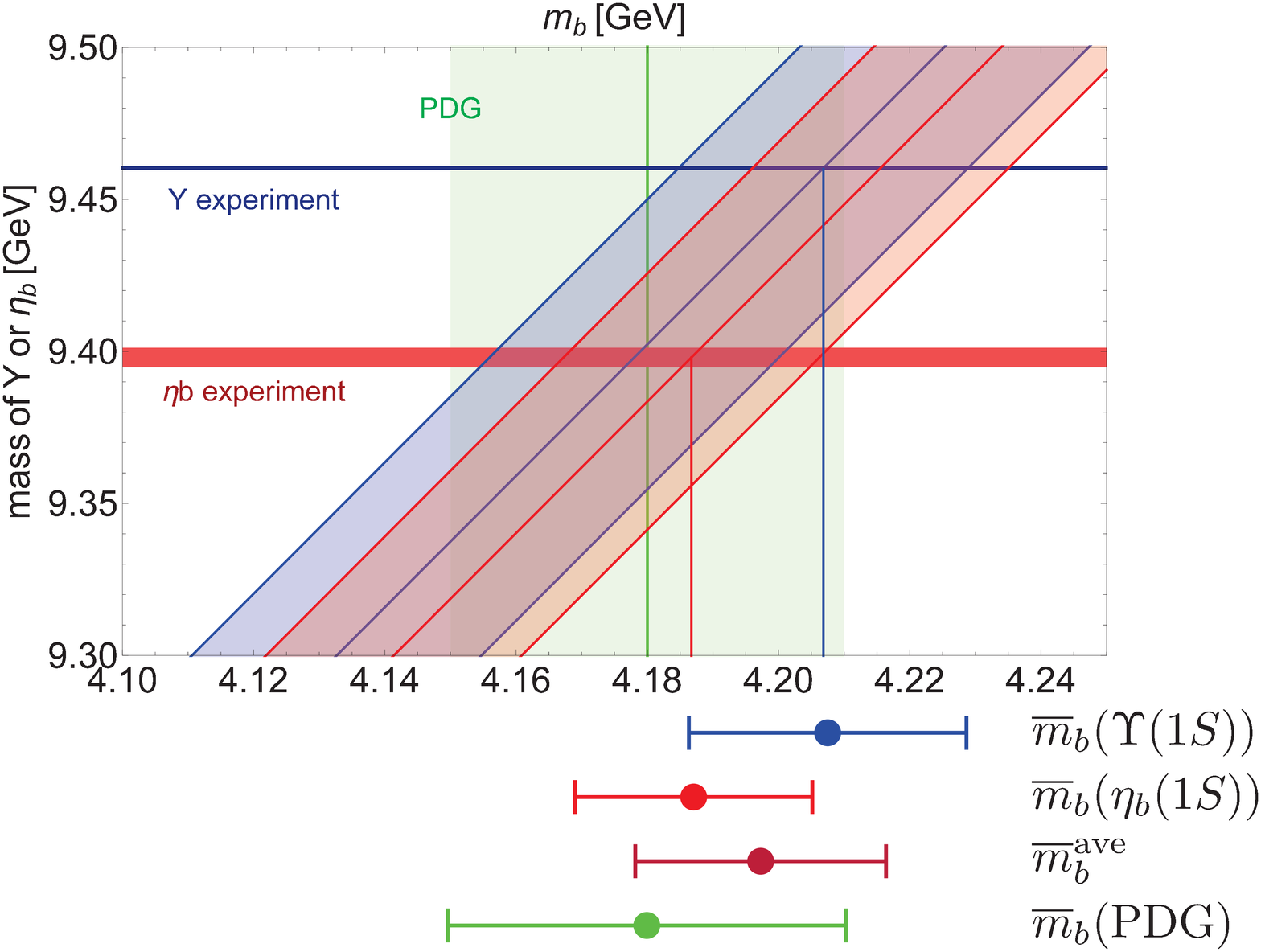}
\end{tabular}
\end{center}
\vspace*{-5mm}
\caption{\small
Determination of $\overline{\rm MS}$ masses
$\overline m_c$ and $\overline m_b$ \cite{Kiyo:2015ufa}.
Horizontal (vertical) axes represent
$\overline m_{c,b}$ 
(mass of charmonium/bottomonium $1S$ state).
Horizontal narrow bands show the experimental data with errors.
Diagonal bands show the perturbative QCD predictions with
errors as functions of $\overline m_{c,b}$.
Determined $\overline m_{c,b}$ with error bars are shown below the plot.
For comparison, the PDG
values are also shown.
}
\label{fig:charm}
\end{figure}

We show
in Fig.~\ref{fig:charm} a (preliminary) prediction
for the bottomonium spectrum in perturbative QCD (red points)
compared with the experimental data (black points).
\begin{figure}
\begin{center}
\includegraphics[width=9cm]{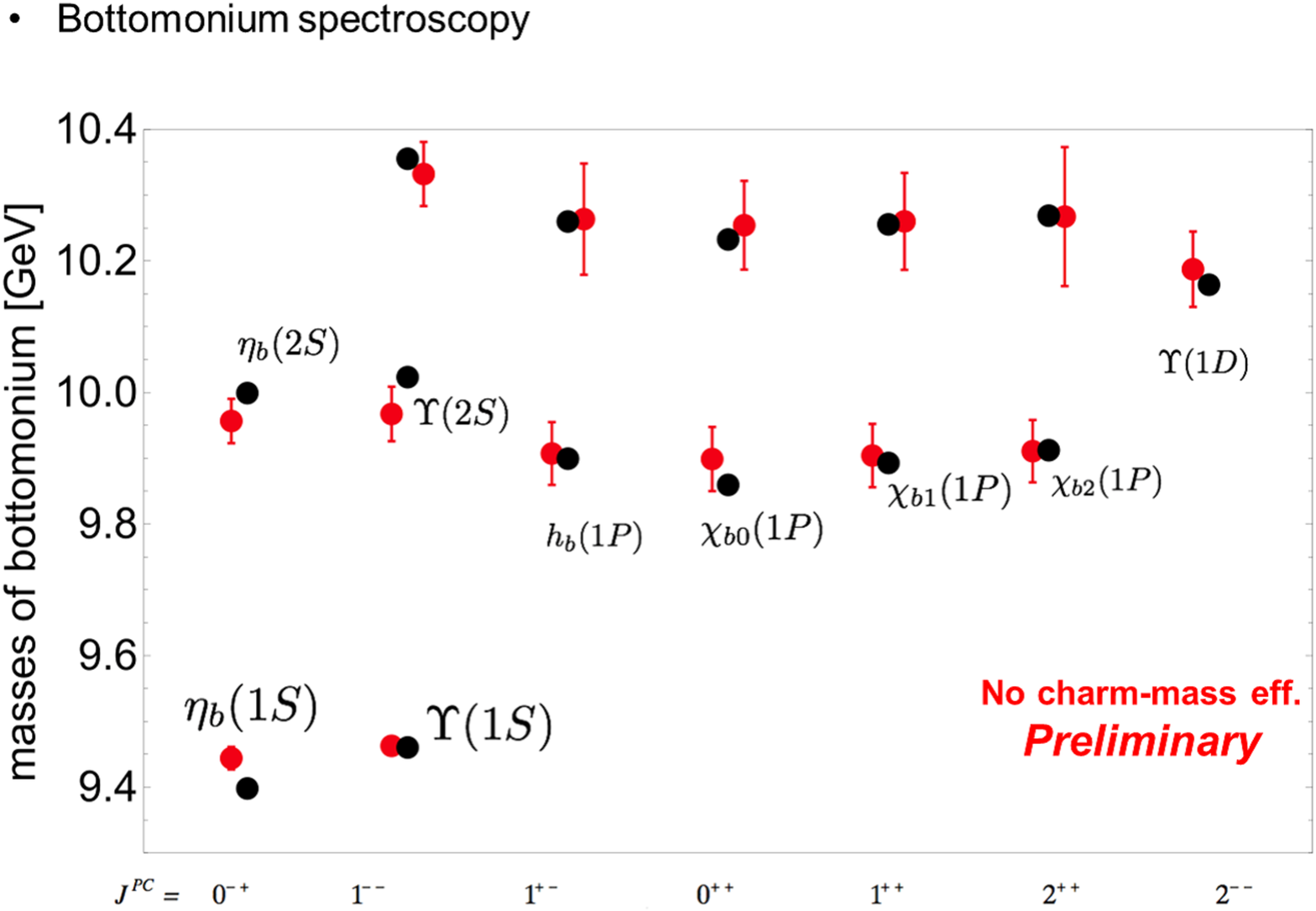}
\end{center}
\vspace*{-.5cm}
\caption{\small
Comparison of the prediction of bottomonium spectrum by perturbative
QCD and the experimental data.
The prediction does not include non-zero charm-mass effects inside loops.
(This figure was made by G.~Mishima.)
\label{BottomoniumSpect}}
\end{figure}
Experimental errors are smaller than the black points,
while theoretical error estimates are shown by red bars.
The bottom quark $\overline{\rm MS}$ mass $\overline{m}_b$ is fixed on the
$\Upsilon(1S)$ state and the value of $\alpha_s(M_Z)$ to
the PDG value.
The prediction \cite{Kiyo:2013aea} does not include non-zero charm-mass effects inside loops
in this figure.
The prediction is in reasonable agreement with the experimental data,
with essentially only $\overline{m}_b$ as the adjustable input parameter.

We can compare the perturbative QCD predictions and the experimental
data for the $1S$ energy levels and determine\footnote{
In this analysis we have included non-zero charm-mass effects inside loops
for the bottomonium $1S$ states.
} the 
$\overline{\rm MS}$ masses of charm and bottom quarks,
$\overline m_c$ and $\overline m_b$.
This is shown in Fig.~\ref{fig:charm}.
From the average of the values determined from the vector and scalar
$1S$ states, we obtain \cite{Kiyo:2015ufa}
$
\overline{m}_c=1246\pm 2 (d_3) \pm 4 (\alpha_s) \pm 23 ({\rm h.o.} )~{\rm MeV}
$
and
$
\overline{m}_b=4197\pm 2 (d_3) \pm 6 (\alpha_s) \pm 20 ({\rm h.o.} )\pm 5  (m_c)~ {\rm MeV}
$,
which agree with the current Particle Data Group values.
The error estimates are based on standard methods for
estimating perturbative
uncertainties.
Thus, the agreement suggests that non-perturbative corrections to these systems
are (at most) comparable in size with the perturbative uncertainties,
which is also consistent with renormalon analyses.

\section{Challenge: Analytic evaluation of $a_3$ (3-loop QCD potential)}

\begin{figure}
\begin{center}
\includegraphics[width=5cm]{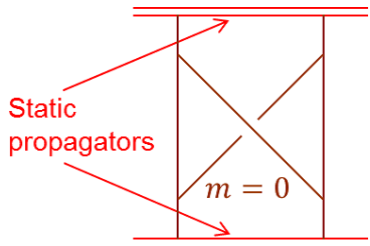}
\end{center}
\vspace*{-.5cm}
\caption{\small
Diagram whose expansion coefficient is not known analytically.
Single lines denote relativistic massless propagators, while double
lines represent static propagators.
\label{NonPlanerDiagram}}
\end{figure}

Currently the three-loop correction to the static QCD potential is
known only numerically \cite{Anzai:2009tm}.
There remain three necessary expansion coefficients in $\epsilon$ of the master integrals
whose analytical values are still unknown \cite{Smirnov:2010zc}.
One of them is an expansion coefficient of the integral given by
the diagram in Fig.~\ref{NonPlanerDiagram}.
Let us explain the status of this coefficient.
This expansion coefficient is reduced to the form
\bea
M_{40}^{(1)}
= \sum_k r_k \,\, Z\!\left(\infty;\{b_{j}^{(k)}\};\{\lambda_{j}^{(k)}\}\right) 
+ C_0 \,,
\eea
namely, as a sum of generalized multiple zeta values
(MZVs) (with rational number coefficients $r_k$)
and an unknown constant $C_0$.
Here, generalized MZV is defined as
\bea
Z\left(\infty;\{b_{j}\};\{\lambda_{j}\}\right) =
\!\!\!\!\!
\sum_{n_{1}>n_{2}>\cdots>n_{N}>0}\frac{\lambda_{1}^{n_{1}}\lambda_{2}^{n_{2}}\cdots\lambda_{N}^{n_{N}}}{n_{1}^{b_{1}}n_{2}^{b_{2}}\cdots n_{N}^{b_{N}}},
~~~~~~~~
b_j\in \mathbb{N},~ 
\lambda_j \in \mathbb{C}.
\label{generalizedMZV}
\eea

The following integral includes the constant $C_0$
whose value has not been
expressed in terms of MZVs up to now:
\bea
&&
\int^\Lambda_0 dx\,\frac{1}{\sqrt{x}\sqrt{1-4x}}\int_{1/4}^xdy\,\frac{1+2y}
{y\sqrt{1-y}\sqrt{1-4y}}\int^y_1dz\,\frac{{\rm Arctan}\,z}{z\sqrt{1-z}}
\nonumber
\\ &&
~~
= C_2 \, \log^2\Lambda + 
C_1 \, \log\Lambda + C_0 + {\cal O}\Bigl(\frac{1}{\Lambda}\log^n \Lambda\Bigr)
~~~~~~
\mbox{as~~$\Lambda\to\infty$}
.
\eea
We consider the limit $\Lambda\to \infty$ and extract the coefficients
$C_0, C_1, C_2$ as above.
$C_2$ and $C_1$ can be expressed by MZVs, which is a
result of
a non-trivial analysis.\footnote{
One can show this using the Cauchy theorem and pseudo-elliptic 
integrals.
}
The constant $C_0$ has not been expressed by MZVs and its nature is still
unknown.
Similar nested integrals with square roots have also been investigated
in this workshop.
We would like to
invite our colleagues to reveal the nature of $C_0$.

\section{Summary}

As demonstrated in this talk,
non-relativistic bound state theory for QCD (also for QED) has become fairly mature 
and amenable to a textbook-level understanding and computation.
A number of recent 
computations use pNRQCD EFT to organize perturbative expansions systematically.

In the computation of NNNLO heavy quarkonium
spectrum
we applied a new technology for evaluating multiple sums
(which may be useful in other applications also)
and performed all computation arithmetically in contrast to the previous
approach which involved diagrammatic analyses.
We also note that after many years of endeavor the computation of
the NNNLO corrections to
the quark pair-production cross section near threshold in
$e^+e^-$ collisions has recently been completed \cite{Beneke:2015kwa}.
One example of the remaining theoretical challenges is an analytic
evaluation of the three-loop correction to the static QCD potential.
It involves evaluation of a new type of nested integral with
square roots.

We demonstrated some physics applications of the NNNLO heavy quarkonium
spectrum.
The prediction shows stability and convergence expected
for a legitimate perturbative prediction.
The prediction of the bottomonium spectrum 
is in reasonable agreement with the
experimental data, in particular by fixing $\alpha_s(M_Z)$ at the PDG value,
with $\overline{m}_b$ as the only adjustable input parameter.
We further determined the 
$\overline{\rm MS}$ masses $\overline{m}_c$ and $\overline{m}_b$ by comparing the
experimental data for $J/\psi(1S),\eta_c(1S)$ and $\Upsilon(1S),\eta_b(1S)$
masses with the
predictions of perturbative QCD.
The obtained values of each mass from the different
spin states
are consistent with each other,
as well as with the current PDG value which
is determined from a wide variety of observables.
These features suggest that non-perturbative corrections to these systems
are comparable in size with the perturbative uncertainties
and therefore under good theoretical control.
Considering the role played by perturbative QCD in the present and
future precision physics, this observation is fairly encouraging.

\vspace{-2mm}
\section*{Acknowledgements}
\vspace{-2mm}
The author is grateful to Y.~Kiyo and G.~Mishima for fruitful collaborations.
This work was supported in part by Grant-in-Aid for
scientific research No.\ 26400238 from
MEXT, Japan.

\end{document}